\documentclass[fleqn,usenatbib]{mnras}

\usepackage{newtxtext,newtxmath}

\usepackage[T1]{fontenc}

\DeclareRobustCommand{\VAN}[3]{#2}
\let\VANthebibliography\thebibliography
\def\thebibliography{\DeclareRobustCommand{\VAN}[3]{##3}\VANthebibliography}

\usepackage{graphicx}	
\usepackage{amsmath}	
\usepackage[dvipsnames]{xcolor}

\title[Probing HeEoR through $^{3}\mathrm{He}^{+}$ transition line]{Probing Helium Reionization Through the $^{3}\mathrm{He}^{+}$ Hyperfine Transition Line}

\author[Basu et al.]{
Arghyadeep Basu,$^{1,2,3}$\thanks{E-mail: basu.arghyadeep@yahoo.in}
Benedetta Ciardi,$^{1}$
Enrico Garaldi$^{4}$
\\
$^{1}$Max-planck-Institut f$\ddot{u}$r Astrophysik, Karl-Schwarzschild-Strasse 1, D-85741, Garching, Germany\\
$^{2}$Ludwig-Maximilians-Universität München (LMU), Geschwister-Scholl-Platz 1, 80539 München, Germany\\
$^{3}$Univ Lyon, Univ Lyon1, Ens de Lyon, CNRS, CRAL UMR5574, F-69230, Saint-Genis-Laval, France\\
$^{4}$Kavli IPMU (WPI), UTIAS, The University of Tokyo, Kashiwa, Chiba 277-8583, Japan\\
}

\date{Accepted 2026 September 04. Received 2026 September 04; in original
form 2025 October 15}

\pubyear{\the\year{}}

\begin{document}
\label{firstpage}
\pagerange{\pageref{firstpage}--\pageref{lastpage}}
\maketitle

\begin{abstract} 
We investigate the hyperfine transition of $^{3}\mathrm{He}^{+}$ as a promising probe of the IGM during the final stages of helium reionization. Utilising the most recent helium reionization simulation, we generate three-dimensional maps of the 3.5cm ($8.67$\,GHz) differential brightness temperature and analyze its evolution. Our `fiducial' results show that the volume-averaged brightness temperature declines rapidly from $\sim 10^{-2} \mu$K (in the `optimistic' approach, i.e. $T_{\rm s} = T_{\rm gas}$, $\sim 1 \mu$K) at $z = 3$ to $\sim 8 \times 10^{-5} \mu$K ($\sim 2.5 \times 10^{-3} \mu$K) by $z = 2.3$, tracing the He\,\textsc{ii} to He\,\textsc{iii} transition driven by quasars. The power spectrum of the 3.5cm signal exhibits a scale-dependent evolution, peaking on small scales and declining as reionization progresses. We explore the cross-correlation of the 3.5cm transition line with the distribution of AGNs, which shows a transition from positive to negative correlation as ionized regions grow. 
Although current observational upper limits lie several orders of magnitude above theoretical predictions, future radio arrays such as \texttt{SKA-mid} offer promising prospects. Overall, this study highlights the $^{3}\mathrm{He}^{+}$ hyperfine transition as a sensitive tracer of the thermal and ionization history of the IGM during helium reionization. The predicted signal is also subject to additional uncertainty from spin temperature modelling, and the true amplitude is expected to lie between the limiting cases explored in this work.

\end{abstract}




\section{Introduction}
At redshift $z = 1100$, hydrogen recombined as the Universe expanded and cooled sufficiently to allow electrons to bind with protons. At earlier times ($z = 6000$), helium experienced its first recombination (He~{\sc iii} $\to$ He~{\sc ii}), i.e. the one pertaining its inner electron, which is bound with an energy of 54.4 eV \citep{Switzer2008b}. As the Universe continued to cool, singly ionized helium (He~{\sc ii}) recombined into neutral helium (He~{\sc i}) at $z= 1800$ \citep{Switzer2008a}. 
With time, the neutral atoms formed during these recombination epochs were gradually reionized by photons emitted from various emerging astrophysical sources.
As the first ionization potential of helium is similar to the one of hydrogen, He~{\sc i} is expected to be ionized into He~{\sc ii} at the same time of hydrogen, i.e. by $z \sim 5.5$ \citep{Fan2006,Becker2015,bosman2018,Bosman2022}. However, the reionization of singly ionized helium (He~{\sc ii} $\to$ He~{\sc iii}) occurs at a later time, requiring the energetic photons emitted by quasars \citep{Compostella2013,upton2016,daloisio2017,Mitra2018,Garaldi2019,Basu2024}. 

Following hydrogen recombination, the IGM comprised approximately $24\%$ helium by number, predominantly in the form of $^4\mathrm{He}$, with a much smaller abundance of $^3\mathrm{He}$—about one part in $10^5$ relative to $^4\mathrm{He}$ \citep{kneller2004}. The $^3\mathrm{He}$ isotope is particularly significant due to its nonzero magnetic dipole moment, which enables hyperfine splitting in the hydrogen-like $^3\mathrm{He}^+$ ion, resulting in a rest-frame transition at 8.67 GHz (3.5cm), analogous to the well-known 21cm line from neutral hydrogen \citep{field1959,madau1997,ciardi2003,furnaletto2006,zaroubi2013,Hassan2016,Mcquinn2018,Ghara2021,Ghara2024,Giri2024,acharya2024}. While much of the effort to explore the high-redshift IGM has focused on the 21cm line, the $^3\mathrm{He}^+$ hyperfine transition offers a unique window into key astrophysical processes in the early Universe \citep{furnaletto2006, bagla2009, McQuinn2009, takeuchi2014, vasiliev2019}. Since the evolution of \textsc{H ii} and He~{\sc ii} during the Epoch of Reionization (EoR) is primarily driven by stellar sources \citep{Marius2018,Marius2020,Basu2025}, the resulting signals at 21cm and 3.5cm are expected to be anti-correlated. Because of that, as proposed by \citet{bagla2009}, beyond tracing the high-$z$ evolution of the helium component in the IGM, the 3.5cm signal could also provide constraints independent from those of the 21cm signal. Additionally, this line can help to constrain the properties of ionizing sources and serve as a powerful mean for detecting quasars \citep{khullar2020}. It also has a potential to probe large-scale filamentary structures as discussed by \citet{takeuchi2014}.

Although the abundance of He~{\sc ii} is lower than the one of \textsc{H ii}, the $^3\mathrm{He}^+$ hyperfine transition has several advantages over the 21cm line, as (i) the foreground contamination is smaller at higher rest-frame frequency, and (ii) its spontaneous decay rate is $\sim 680$ times larger, enhancing the signal strength. However, the detectability of this signal is limited by the sensitivity of most current telescopes. While there are several radio telescopes which are operating in this relevant frequency range, the Square Kilometre Array (\texttt{SKA}; specifically in the Phase 2, using \texttt{SKA-mid}) is the most promising one to detect this signal. 

While \citet{bagla2009} estimated the signal strength using semi-analytic methods, more detailed numerical simulations are necessary to capture the full complexity of He~{\sc ii} evolution, including spatial inhomogeneities, radiative transfer effects, and feedback mechanisms that semi-analytic models may not fully incorporate. \citet{takeuchi2014} provided a precise estimation of the ionization state of $^3\mathrm{He}^+$ and studied the physical processes that influence its spin temperatures over a wide redshift range ($z \sim 0-8$). However, their work did not include radiative transfer effects, which are crucial for a more accurate representation of how ionizing radiation propagates through the IGM. 
The investigation by \citet{khullar2020} addressed some of these limitations by using radiative transfer simulations (\citealt{Marius2020}) to explore the possibility of detecting the $^3\mathrm{He}^+$ signal, considering various ionizing sources, including stars, X-ray binaries, accreting black holes, and shock-heated gas in the interstellar medium. Additionally, they also studied a high-$z$ quasar to investigate its environment (see also \citealt{koki2017}). However, a more detailed investigation at lower redshifts, where helium reionization is ongoing, is crucial for assessing the feasibility of detecting the 3.5cm signal over a wider redshift range. Since the impact of quasars on reionization is redshift-dependent, focusing on this later stage provides new insights into how the evolving He~{\sc ii} ionizing radiation field influences the IGM. Their analysis, however, assumes efficient coupling of the spin temperature to the gas temperature, which can significantly enhance the expected signal. As shown by \citet{McQuinn2009}, the $^3\mathrm{He}^+$ spin temperature is expected to remain tightly coupled to the CMB across most of the IGM at these redshifts, since both radiative and collisional decoupling are inefficient except in extreme environments. This strongly suppresses the observable signal and places stringent constraints on detectability, even in the presence of realistic ionizing backgrounds. During the revision process of this manuscript, more recent work by \citet{Spina2026} revisited the problem with a more consistent treatment of the spin temperature coupling and found that, although the overall signal remains weak, different helium reionization histories can produce distinguishable signatures in the power spectrum, potentially accessible to next-generation instruments.

One the other hand, recent observational work has sought to constrain the prospects of helium hyperfine signal detection. \citet{Trott2024} presented the first limits on the power spectrum  of the $^3\mathrm{He}^+$ hyperfine transition at $z = 3{-}4$ on spatial scales of 30 arcminutes, using 190 hours of archival interferometric data from the Australia Telescope Compact Array (ATCA). Although noise and residual radio frequency interference limited their sensitivity, the study demonstrated the feasibility of detecting the helium signal with current telescopes. It also emphasized the potential of next-generation instruments, such as \texttt{SKA}, which, with their higher sensitivity, could yield cosmologically relevant signals and offer deeper insights into helium reionization.

In this study, we discuss the expected $^3\mathrm{He}^+$ hyperfine transition line  at $z < 5$, estimated from the latest cosmological simulations of helium reionization \citep[][hereafter B24]{Basu2024}. We also perform a quantitative comparison between the simulated $^3\mathrm{He}^+$ signal and the first observational constraints from \citet{Trott2024}. The simulation incorporates recent constraints on the quasar luminosity function (QLF) and successfully reproduces a majority of helium reionization observations, including the He~{\sc ii} Lyman-$\alpha$ forest. Specifically, we introduce the simulation methodology in Section \ref{method}, present the results in Section \ref{results}, and conclude with a discussion in Section \ref{conclusion}. Throughout this work, we adopt a flat $\Lambda \rm{CDM}$ cosmology, consistent with \cite{Planck2016}, using the following parameters: $\Omega_{\rm m} = 0.3089$, $\Omega_{\Lambda} = 0.6911$, $\Omega_{\rm b} = 0.0486$, $h = 0.6774$, $\sigma_{8} = 0.8159$, and $n_{\rm s} = 0.9667$, where the symbols have their usual meanings.

\begin{figure*}
    \includegraphics[width=180mm]{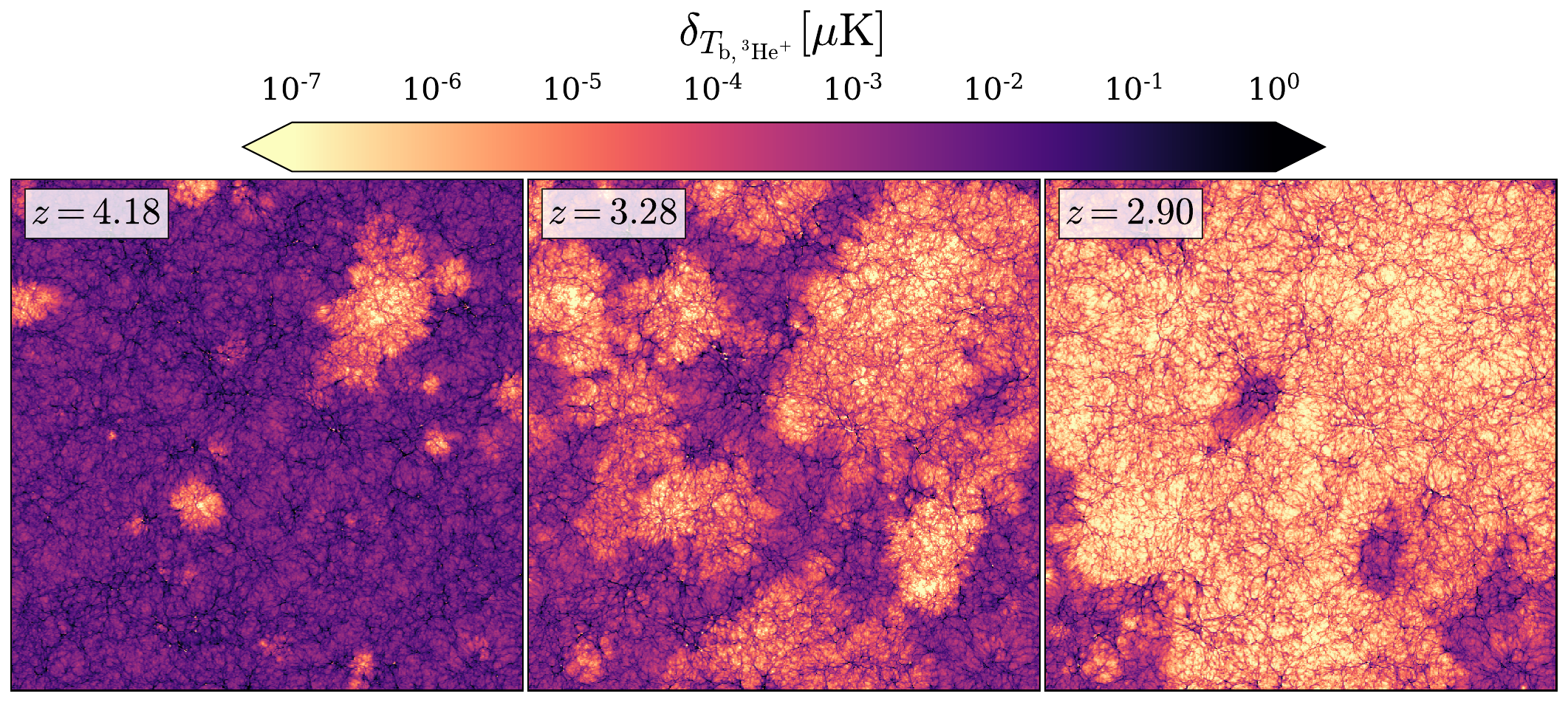}
    \caption{Slice map of $^3\rm{He}^+$ differential brightness temperature across the simulation box at $z=4.18$, 3.28 and 2.90, from left to right.  The maps are 205$\rm{\mathit{h}^{-1} cMpc}$ wide and 400$\rm{\mathit{h}^{-1} ckpc}$ thick. }
    \label{deltatb_map}
\end{figure*}

\section{Methods}
\label{method}

\subsection{Simulation of helium reionization}

In the following, we briefly describe the simulation used to estimate the 3.5cm signal, while we refer the reader to B24 for more details. 

The radiative transfer (RT) simulation has been performed by post-processing the \texttt{TNG300} hydrodynamical simulation, which is part of the \texttt{Illustris TNG} project \citep{volker2018,naiman2018,marinacci2018,pillepich2018,nelson2018}. This simulation was performed using the \texttt{AREPO} code \citep{springel2010}, which solves the idealized magneto-hydrodynamical equations \citep{pakmor2011} governing the non-gravitational interactions of baryonic matter, as well as the gravitational interactions of all matter. \texttt{TNG300} employs the latest \texttt{TNG} galaxy formation model \citep{weinberger2017,pillepich2018}, with star formation implemented by converting gas cells into star particles above a density threshold of $\textit{n}\rm{_{H}} \sim 0.1 \ \rm{cm^{-3}}$, following the Kennicutt-Schmidt relation \citep{springel2003}. The simulation is run in a comoving box of length $L_\mathrm{box} = 205 \,h^{-1} \, \mathrm{cMpc}$, with (initially) $2 \times 2500^{3}$ gas and dark matter particles. The average gas particle mass is $\bar{m}\rm{_{gas}} = 7.44 \times 10^{6} \, M_{\odot}$, while the dark matter particle mass is fixed at $m\rm{_{DM}} = 3.98 \times 10^{7} \, M_{\odot}$. Haloes are identified on-the-fly using a friends-of-friends algorithm with a linking length of $0.2$ times the mean inter-particle separation.  We utilized 19 outputs covering the redshift range $5.53 \geq z \geq 2.32$.

The radiative transfer of ionizing photons through the IGM has been implemented with the \texttt{CRASH} code \citep[e.g.][]{ciardi2001,maselli2003,maselli2009,Maselli2005,partl2011}, which computes self-consistently the evolution of hydrogen and helium ionization states, as well as gas temperature. \texttt{CRASH} employs a Monte Carlo-based ray-tracing scheme, where ionizing radiation and its spatial and temporal variations are represented by multi-frequency photon packets propagating through the simulation volume. The latest version of \texttt{CRASH} includes UV and soft X-ray photons, accounting for X-ray ionization, heating, detailed secondary electron physics \citep{graziani2013,graziani2018}, and dust absorption \citep{glatzle2019,glatzle2022}. For more details, we refer the reader to the original \texttt{CRASH} papers. The RT is performed on grids of gas density and temperature extracted from \texttt{TNG300} snapshots and tracks radiation with energy $h_{\rm P} \nu \in [54.4 \ \mathrm{eV}, 2 \ \mathrm{keV}]$, assuming fully ionized hydrogen (i.e. $x_{\rm HI} = 10^{-4}$) and fully singly ionized helium (i.e. $x_{\rm HeI} = 0$ and $x_{\rm HeIII} = 10^{-4}$). 

As sources of ionizing radiation we consider quasars, which are assigned to halos in the simulation according to a modified abundance-matching approach (see B24 for more details) to reproduce the QLF of \citet{Shen2020}.
The quasar SED is the same adopted in \citet{Marius2018} and \citet{Marius2020}, which is obtained by averaging over 108,104 SEDs observed in the range $0.064 < z < 5.46$ \citep{Krawczyk2013} at $h_{\rm P} \nu <200$~eV, while it follows a power law with index $-1$ at higher energies (see Section 2 of \citealt{Marius2018}).  

\subsection{The 3.5cm signal}
\label{signal_def}

The simulation described in the previous section provides the spatial and temporal distributions of the gas number density, $\rm{\mathit{n}_{gas}}$, and temperature, $\rm{\mathit{T}_{gas}}$, as well as of the fractions of H~{\sc ii}, He~{\sc ii}, and He~{\sc iii} (denoted as $\rm{\mathit{x}_{HII}}$, $\rm{\mathit{x}_{HeII}}$, and $\rm{\mathit{x}_{HeIII}}$, respectively). 
The brightness temperature associated with the hyperfine transition of $^{3}\mathrm{He}^{+}$, $\rm{\mathit{\delta T}_{b,^{3}He^{+}}}$, is evaluated in each cell of the simulated volume as (see eq.~61 of \citealt{furnaletto2006}):
\begin{align}
    \rm{\mathit{\delta T}_{b,^{3}He^{+}}} = & \ 0.5106 \ \rm{\mathit{x}_{HeII} \ (1+\delta) \ (1 - \frac{\mathit{T}_{CMB}}{\mathit{T}_{s}})} \nonumber \\
    & \times \ \rm{(\frac{[^{3}He/H]}{10^{-5}}) \ (\frac{\Omega_{b}\mathit{h}^{2}}{0.0223}) \sqrt{\frac{\Omega_{m}}{0.24}} \ (1+\mathit{z})^{1/2} \ \mu K}, 
    \label{deltatb_formula}
\end{align}
where $\delta = (\mathit{n}_{\mathrm{gas}} - \bar{\mathit{n}}_{\mathrm{gas}})/\bar{\mathit{n}}_{\mathrm{gas}}$ is the gas overdensity, with $\bar{\mathit{n}}_{\mathrm{gas}}$  mean gas number density within the simulation box. $\mathit{T}_{\mathrm{CMB}} = 2.725(1+\mathit{z}) \, \mathrm{K}$ denotes the CMB temperature at redshift $z$, $\rm{\mathit{T}_{s}}$ is the spin temperature, and the relative abundance of $^{3}\mathrm{He}$ to hydrogen, $[\mathrm{^{3}He/H}] = 10^{-5}$, is determined by Big Bang nucleosynthesis. 
We adopt the common assumptions $\frac{\mathit{H}(z)}{(1+\mathit{z})(dv/dr)} \sim 1$.
For the spin temperature, we account for coupling to the CMB, collisional excitation and de-excitation, and Ly$\alpha$ scattering, adopting the equilibrium expression $T_{\rm s}^{-1} = \left( T_{\mathrm{CMB}}^{-1} + x_{\alpha} T_{\alpha}^{-1} + x_{\rm c} T_{\rm gas}^{-1} \right) / \left( 1 + x_{\alpha} + x_{\rm c} \right)$, where $T_{\rm gas}$ is the gas temperature as defined before, $T_{\rm \alpha}$ is the color temperature, and $x_c$ and $x_{\alpha}$ are the collisional and radiative coupling coefficients, respectively (defined as our fiducial' approach). The collisional coupling coefficient $x_{\rm c}$ is computed using the prescription of \citet{McQuinn2009}. Since local $^3\mathrm{He}^{+}$ Ly$\alpha$ photons are expected to be too scarce to significantly influence $T_{\rm s}$  \citep[see][]{McQuinn2009}, and their effect is further reduced by competing absorption from $^4\mathrm{He}^{+}$, we neglect radiative coupling and set $x_{\alpha}=0$. We note, though, that with this assumption the results should be considered as lower limits for the signal. On the other hand, an upper limit would be obtained by assuming full coupling with gas temperature, i.e. $T_{\rm s} = T_{\rm gas}$ as done in \citet{khullar2020} (the `optimistic' approach). We discuss how this assumption impacts our results in Appendix.\ref{compared_PS_with_Ts_treatment} as well as in Figure \ref{ps}.

The power spectrum (PS) of the differential brightness temperature is defined as:
\begin{align}
    \label{ps-equ}
    \rm{\mathit{P}_{^{3}He^{+}}(k) = \langle \ \mathit{\delta T}_{b,^{3}He^{+}} (k) \mathit{\delta T}_{b,^{3}He^{+}}(k)^{*} \rangle},
\end{align}
where $\rm{\mathit{\delta T}_{b,^{3}He^{+}} (k)}$ is the Fourier transform of the differential brightness temperature field, and $\rm{\mathit{\delta T}_{b,^{3}He^{+}} (k)^{*}}$ is its complex conjugate. 
In the following, we will express the results in terms of the dimensionless power spectrum, defined as:
\begin{align}
    \rm{\Delta_{^{3}He^{+}}^{2} = \frac{\mathit{k}^{3}}{2 \pi^{2}} \mathit{P}_{^{3}He^{+}}(k)}.
\end{align}

\begin{figure}
    \includegraphics[width=\columnwidth]{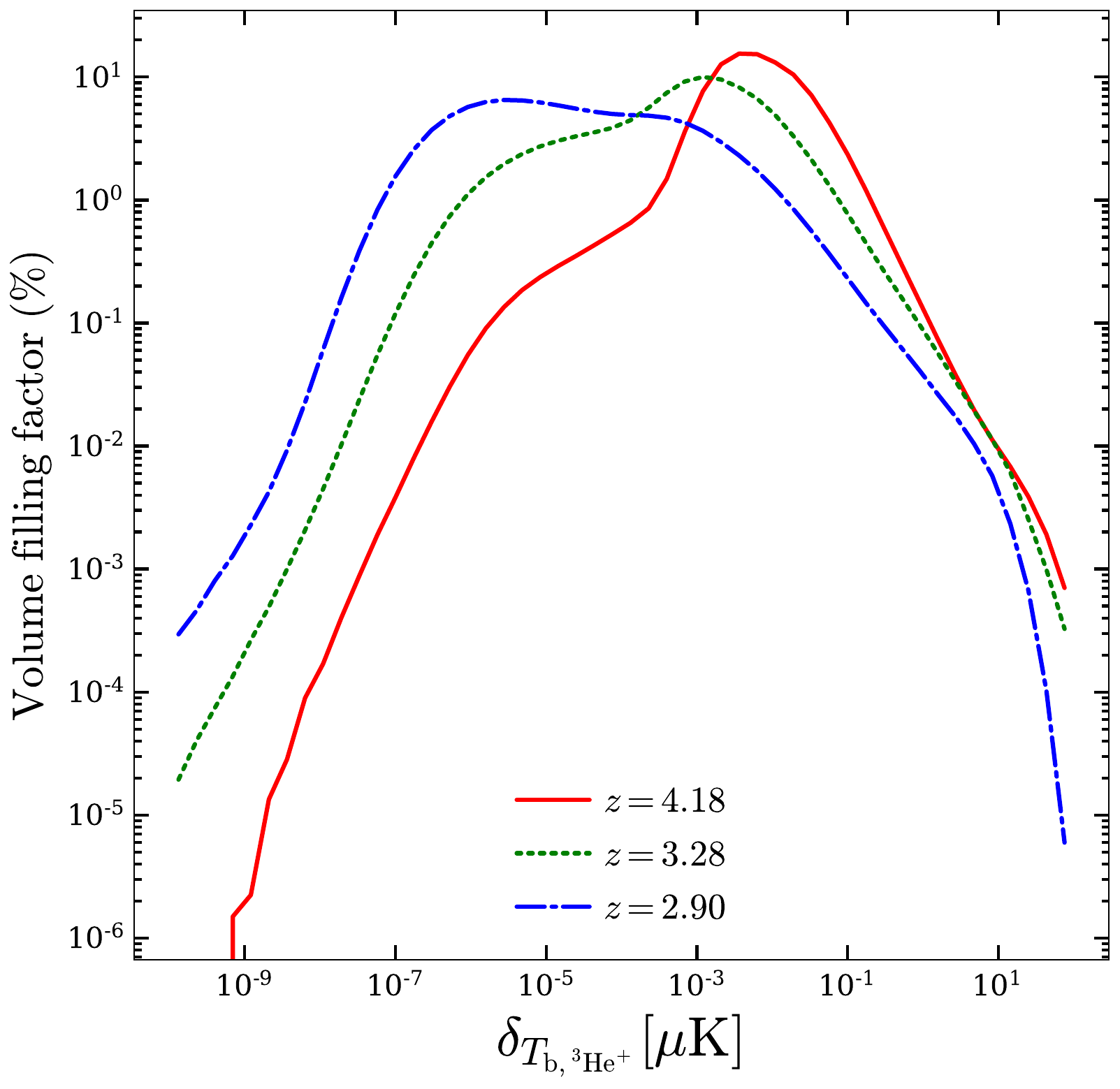}
    \caption{Volume filling factor of differential brightness temperature of the hyperfine transition of $^{3}\rm{He}^{+}$ at $z=4.18$ (solid red curve), 3.28 (dotted green) and 2.90 (dash-dotted blue). }
    \label{deltatb_dist}
\end{figure}

\section{Results}
\label{results}

In this section, we discuss the features and potential observables associated to the 3.5cm transition line under the `fiducial' treatment (see Section~\ref{signal_def}) extracted from the helium reionization simulations introduced in B24.

\subsection{Overview}

In Figure \ref{deltatb_map} we present maps of $\rm{\mathit{\delta T}_{b,^{3}He^{+}}}$ for a slice of the simulation box at   $z = 4.18$, $3.28$, and $2.90$, corresponding to the redshifts
of the observations by \citet{Trott2024}. At $z = 4.18$, most of the helium in the simulation volume is in the form of He\,\textsc{ii}, resulting in  values of $\rm{\mathit{\delta T}_{b,^{3}He^{+}}}$ typically exceeding $0.01\,\mu\mathrm{K}$. However, in regions that have been fully ionized to He\,\textsc{iii} by the ionizing radiation emitted by quasars, $\rm{\mathit{\delta T}_{b,^{3}He^{+}} (k)}$ drops to $ 10^{-6}\,\mu\mathrm{K}$. At $z = 3.28$ more gas has low values of $\rm{\mathit{\delta T}_{b,^{3}He^{+}}}$, reflecting the expansion of He\,\textsc{iii} regions throughout the volume. By $z = 2.90$, nearly the entire slice of the simulation box exhibits low $\rm{\mathit{\delta T}_{b,^{3}He^{+}}}$ values, marking the near-completion of helium reionization. 

For a more quantitative analysis, Figure~\ref{deltatb_dist} shows the volume filling factor of $\rm{\mathit{\delta T}_{b,^{3}He^{+}}}$ at the same redshifts as the slice maps in Figure~\ref{deltatb_map}. At all redshifts, the volume filling factor shows a double-peaked structure. The peak at higher $\rm{\mathit{\delta T}_{b,^{3}He^{+}}}$ values corresponds to regions where helium is still singly ionized, while the lower peak is linked to areas where helium has been fully ionized to He\,\textsc{iii}. At $z = 4.18$, the fraction is dominated by a strong peak around $0.01\,\mu\mathrm{K}$, which is over 2 orders of magnitude higher than the `shoulder' like feature at $\delta T_{\mathrm{b},^{3}\mathrm{He}^{+}} \sim 10^{-6}\,\mu\mathrm{K}$. 
By $z = 3.28$, the volume filling factor at $\delta T_{\mathrm{b},^{3}\mathrm{He}^{+}} \lesssim 10^{-3}\,\mu\mathrm{K}$ increases significantly due to ongoing helium reionization, while the contribution from higher emission regions declines. At $z = 2.90$, the peak prominence shifts entirely to lower $\rm{\mathit{\delta T}_{b,^{3}He^{+}}}$ values, with the dominant peak centered around $4 \times 10^{-7}\,\mu\mathrm{K}$, marking the completion of helium reionization.

\begin{figure}
    \includegraphics[width=\columnwidth]{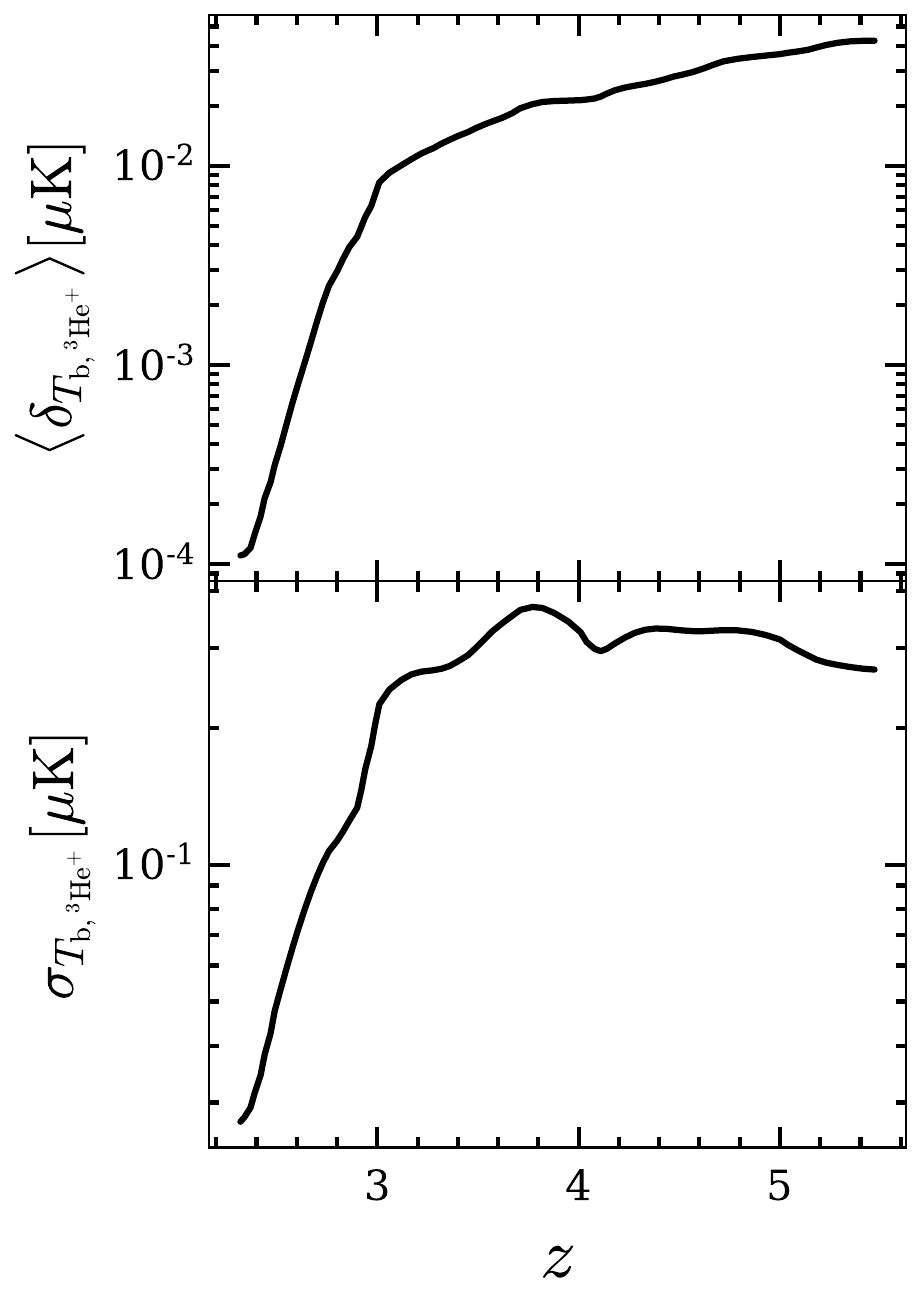}
    \caption{Redshift evolution of the volume averaged differential brightness temperature (\textit{top panel}) and standard deviation (\textit{bottom}) of the hyperfine transition of $^{3}\rm{He}^{+}$.}
    \label{3he_evolution}
\end{figure}

In the top panel of Figure \ref{3he_evolution} we show the volume-averaged differential brightness temperature, i.e. $\langle \delta T_{\mathrm{b},^{3}\mathrm{He}^{+}} \rangle$. Throughout the entire redshift range $\langle \delta T_{\mathrm{b},^{3}\mathrm{He}^{+}} \rangle$ remains positive because $T{\rm _S}$ is always higher than the CMB temperature (see Equation \ref{deltatb_formula}). 
The redshift evolution of $\langle \delta T_{\mathrm{b},^{3}\mathrm{He}^{+}} \rangle$ exhibits a gradual, marginal decline until $z=3$, when about $\sim$80\% of He~{\sc ii} has been ionized. The trend qualitatively follows the one of the He~{\sc ii} fraction (see Figure 3 in B24). At lower redshift (below $z=3$) the average signal starts to decline rapidly, starting from a value of $10^{-2} \mu$K, reaching about $ 8 \times 10^{-5}\mu$K by $z= 2.3$. This drop is linked to the completion of helium reionization. For completeness, we also report the results obtained under the `optimistic' assumption (as discussed in Section~\ref{signal_def}). In this case, the average signal starts from $\sim 1,\mu$K at high redshift and decreases rapidly, reaching $\sim 2.5 \times 10^{-3},\mu$K by $z = 2.3$.

To track how fluctuations in the global 3.5cm signal evolve over time, in the bottom panel of Figure \ref{3he_evolution} we examine the evolution of the standard deviation of $\delta T_{\rm b,^{3}\rm{He}^{+}}$, i.e. $\sigma_{\delta T_{\mathrm{b},^{3}\mathrm{He}^{+}}}$. 
Aside from a similar early time increase associated with the starting of He~{\sc ii} reionization, as also seen for $\delta T_{\rm b,^{3}\mathrm{He}^{+}}$, the fluctuations remain largely consistent (with minor short time-scale variations) up to $z=3$.
By this time, about 80\% of He~{\sc ii} in the simulation volume has been fully ionized into He~{\sc iii}. After $z= 3$, the fluctuations start to decrease because the reionization process has smoothed out most variations in the abundance of He~{\sc ii} in the IGM. This transition occurs at a redshift when the temperature of the IGM at mean density begins to saturate, as observed in B24. Once reionization is complete, the remaining fluctuations are induced by fluctuations in the gas density.

In the following sections, we will focus on potentially observable quantities related to the 3.5cm transition line and their trends with redshifts.

\subsection{Power spectra of 3.5cm signal}

\begin{figure}
    \includegraphics[width=\columnwidth]{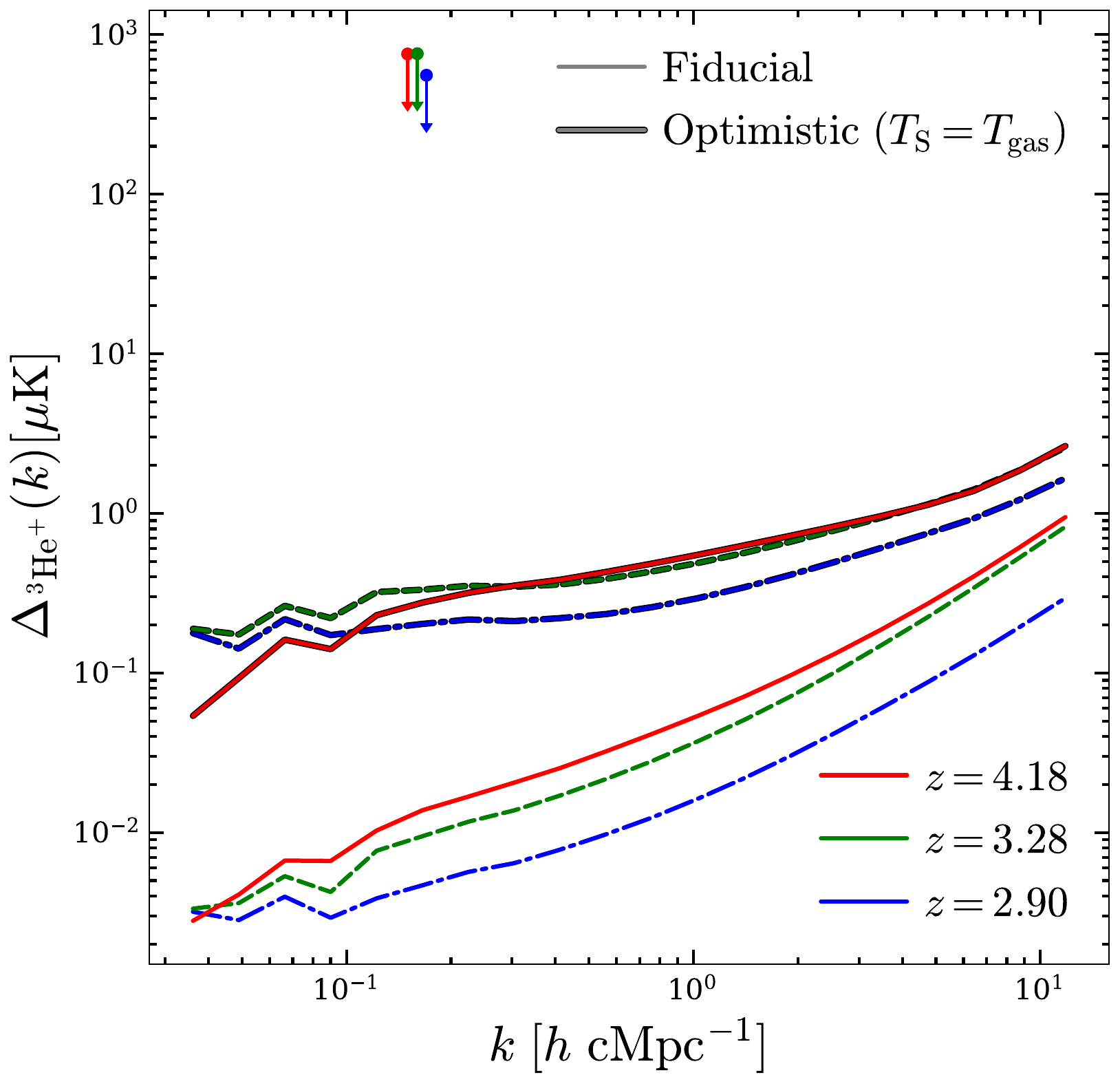}
    \caption{Power spectra of the 3.5cm signal at $z=4.18$ (solid red), $z=3.28$ (dotted green), and $z=2.90$ (dash-dotted blue). Results obtained using the fiducial spin temperature treatment are shown by the smooth coloured curves, while those corresponding to the optimistic limit are indicated by the same curves with black outlines. Observational upper limits from \citet{Trott2024} at the corresponding redshifts are shown as circles. The true result is expected to lie between these two limiting cases.}
    \label{ps}
\end{figure}

\begin{figure}
    \includegraphics[width=\columnwidth]{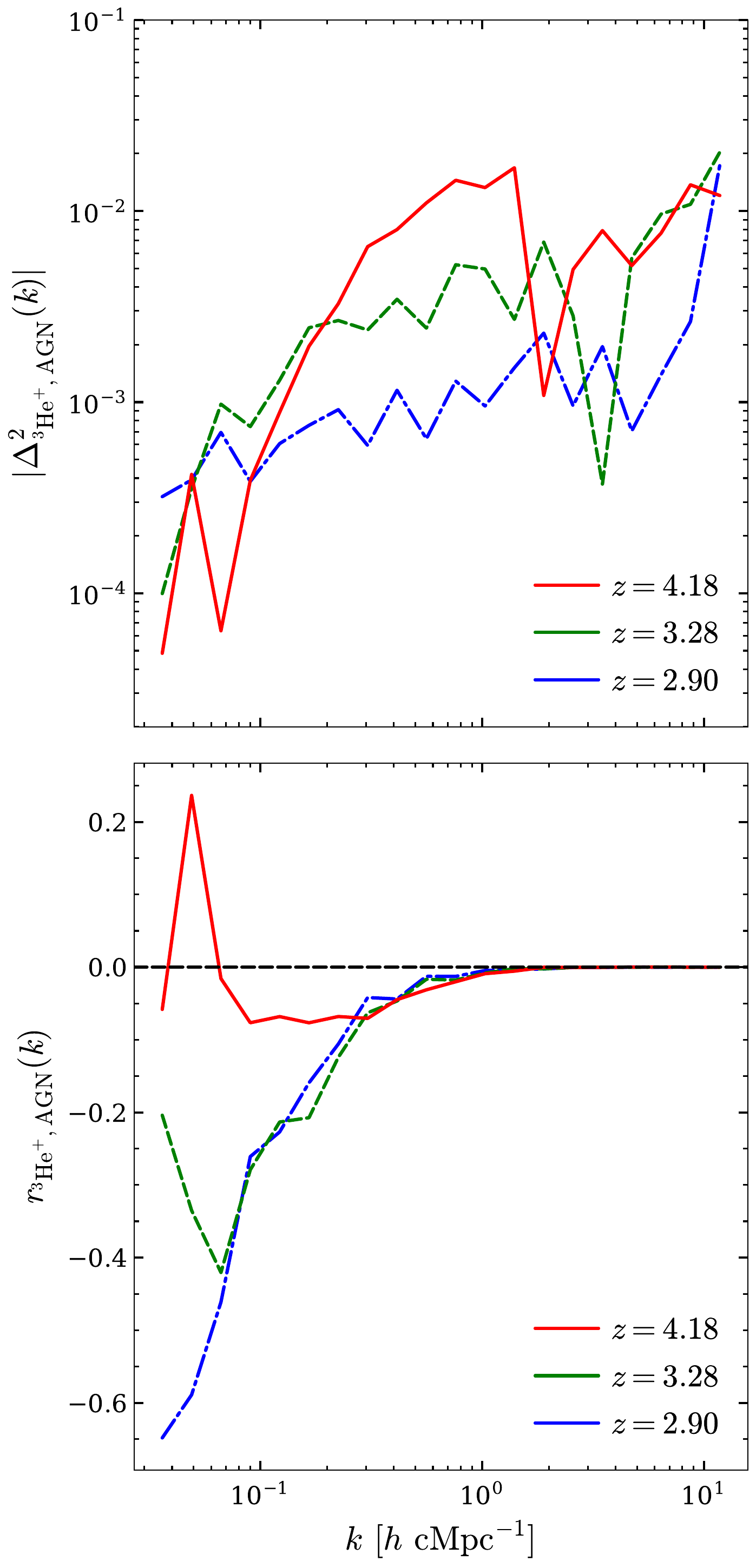}
    \caption{\textit{Top panel:} Simulated 3.5cm-AGN dimensionless cross-power spectra at $z=4.18$ (solid red curve), 3.28 (dotted green) and 2.90 (dash-dotted blue). \textit{Bottom:}  Cross-correlation coefficient between these two fields at the same redshifts. The black dotted curve denotes no correlation  as a reference.}
    \label{he3_qso_crossPS}
\end{figure}


Similarly to the 21cm line, the power spectrum of the signal is expected to be the first detectable quantity. Thus, in Figure \ref{ps} we present the dimensionless power spectra as $\rm{\Delta_{^{3}He^{+}}}(\mathit{k})$. All curves increase towards higher $k$ values, indicating stronger fluctuations at smaller spatial scales. At $z\gtrsim 5$, $\rm{\Delta_{^{3}He^{+}}}(\mathit{k})$ is mainly dominated by the over-density of gas matter and the inhomogeneous temperatures from the hydrodynamic simulations.
As reionization proceeds, the impact of quasars emerges in a scale dependent way. 
From $z = 4.18$ to $3.28$, the amplitude of the fluctuations at $\rm{k \gtrsim 0.2 \, \mathit{h} \, Mpc^{-1}}$ (corresponding to spatial scales of $\leq 50 \, \rm{\mathit{h}^{-1} \, Mpc}$) remains largely unchanged, while it drops by $z=2.90$. Essentially, the quasars radiation increases the gas temperature in their vicinity, slightly reducing the fluctuations of $\rm{\mathit{\delta T}_{b,^{3}\rm{He}}}$, as well as the amplitude of  $\rm{\Delta_{^{3}He^{+}}}(\mathit{k})$. On the larger spatial scales, the effect of the inhomogeneous heating becomes more pronounced (see also, \citealt{pritchard2007}), increasing the amplitude of $\rm{\Delta_{^{3}He^{+}}}(\mathit{k})$ from $z = 4.18$ to $3.28$ (when, $\rm{\mathit{x}_{HeII}}$ drops from $85\%$ to less than $50\%$, see also Figure 3 in B24). By $z = 2.90$, $\rm{\Delta_{^{3}He^{+}}}(\mathit{k})$ decreases further since the majority of the simulation volume is fully ionized, and fluctuations are reduced at all scales.

Figure \ref{ps} further compares our fiducial spin temperature treatment with the optimistic limiting case of fully coupled spin and gas temperatures ($T_{\rm S}=T_{\rm K}$). The fully coupled approximation systematically increases the power spectrum amplitude at all redshifts and scales, with the largest enhancement occurring on intermediate and large scales where temperature fluctuations are spatially coherent and closely trace the gas temperature field. In contrast, the self-consistent fiducial model introduces additional spatial variations in the coupling efficiency, partially decorrelating the spin and gas temperatures and reducing the brightness-temperature contrast. The relative difference between the two models decreases towards smaller scales, where local density fluctuations dominate the signal and the details of spin temperature coupling become less important. Despite these differences, the qualitative redshift evolution remains unchanged in both models. 

In the Figure \ref{ps}, we also include the first observational upper limits on the 3.5cm brightness temperature fluctuation from \citet{Trott2024}.These upper limits exceed our fiducial theoretical predictions by approximately 5-6 orders of magnitude. Even in the optimistic limit, the predicted signal remains about 4 orders of magnitude below the observational constraints. This is primarily attributed to the adopted simple model of foreground emission in \citet{Trott2024} and to the sensitivity limitations of the telescope.

\subsection{Cross-correlation between the 3.5cm signal and AGNs}

As initially discussed by \citet{lidz2009}, cross-correlation between galaxies and 21cm emission promises to be an excellent probe of the EoR, as it is sensitive to the size and filling factor of H~{\sc ii} regions, and it would alleviate the effect of observational systematics. 
This topic has been extensively investigated in recent years (for e.g. \citealt{wiersma2013,vrbanec2016,hutter2018,vrbanec2020,hutter2023,laplante2023,hartman2025}). 
A similar approach could be applied to the 3.5cm signal with AGNs, as it could provide additional information on the timing and morphology of helium reionization. 
To explore this, we compute the cross-power spectrum between the differential brightness temperature of the \(^3\mathrm{He}^+\) signal and the spatial distribution of AGNs \footnote{We use the terminology `AGN' here rather than `QSO', as our simulated sources extend to faint luminosities.}
 in our simulations. The AGN density field can be characterized by the density contrast \(\delta_{\text{AGN}}(\mathbf{x}) = (n_{\text{AGN}}(\mathbf{x})/{\langle n_{\text{AGN}} \rangle}) - 1\),  where \(n_{\text{AGN}}(\mathbf{x})\) is the local number density of AGNs, and \(\langle n_{\text{AGN}} \rangle\) is its spatial average \footnote{Note that, we calculate this quantity on a cell-by-cell basis over the simulation grid.}. Following the methodology of \citet{lidz2009}, we calculate the cross-power spectrum as a function of wave number \(k\). The top panel of Figure~\ref{he3_qso_crossPS} shows the resulting cross-power spectrum, while the bottom panel presents the corresponding cross-correlation coefficient, defined as \(r_{\rm ^{3}He^{+},\text{AGN}}(k) = P_{\rm ^{3}He^{+},\text{AGN}}(k)/{\sqrt{P_{\rm ^{3}He^{+}}(k) P_{\text{AGN}}(k)}}\), where \(P_{\rm ^{3}He^{+},\text{AGN}}(k)\) is the cross-power spectrum, and \(P_{\rm ^{3}He^{+}}(k)\) and \(P_{\text{AGN}}(k)\) are the auto-power spectra of the \(^3\mathrm{He}^+\) signal and AGN density field, respectively. 

At $z = 4.18$, the cross-power spectrum peaks at $k \sim 1 \, \mathrm{\mathit{h}\, cMpc^{-1}}$, while the cross-correlation coefficient indicates a positive correlation between AGNs and the 3.5cm signal on the largest spatial scales ($k \lesssim 0.1 \, \mathrm{\mathit{h}\, cMpc^{-1}}$). At this stage, AGNs are still rare and have only just begun ionizing their local environments. Since they preferentially reside in overdense regions, which contain more matter and thus more He\,\textsc{ii}, these regions exhibit stronger 3.5cm emission prior to being ionized. The resulting positive correlation on large scales does not arise from direct AGN emission, but rather from the fact that both the AGNs and the 3.5cm signal trace the same large-scale overdense regions in the IGM that are rich in He\,\textsc{ii}. A subdominant contribution to the positive correlation peak at $z \sim 4$ may also arise from a numerical artifact of our source modelling scheme, although this effect does not dominate the signal.

As ionized bubbles begin to grow around these AGNs, the 3.5cm emission in their immediate surroundings starts to decrease. This lowers the amplitude of the cross-power spectrum at intermediate and small scales (larger $k$), even though the cross-correlation coefficient at these scales gradually transitions toward negative values. Physically, the anti-correlation strengthens because regions hosting AGNs now correspond to He\,\textsc{iii} bubbles, where the 3.5cm signal is suppressed, while the emission is maintained in the more distant, mostly underdense regions that lack AGNs. By $z = 3.28$, the cross-power spectrum amplitude has dropped, and the peak gets smoothened toward larger spatial scales, reflecting the typical size of He\,\textsc{iii} bubbles. At this stage, the cross-correlation coefficient is negative up to $k \sim 1 \, \mathrm{\mathit{h}\, cMpc^{-1}}$. By $z = 2.90$, when most of the volume is ionized, the cross-power spectrum continues to decline at all scales, and the anti-correlation remains to lower $k$. On the smallest scales (high $k$), the cross-correlation coefficient approaches zero, as the remaining 3.5cm emission becomes uncorrelated with the sparse AGN distribution. Throughout this evolution, the curves remain noisier than standard 21cm--galaxy cross-power spectra due to the much lower AGN number density. Nevertheless, these trends demonstrate that the 3.5cm--AGN cross-power spectrum provides a sensitive probe of He\,\textsc{iii} bubble growth and the morphology of the helium reionization process.

\section{Discussion and Conclusions}
\label{conclusion}
In this study, we investigated the $^{3}\mathrm{He}^{+}$ hyperfine transition line at 8.67\,GHz (3.5cm) as a promising probe of helium reionization in the intergalactic medium (IGM) at $z < 5$. Earlier studies have explored the potential of the $^{3}\mathrm{He}^{+}$ transition under different physical conditions and modelling approaches. \citet{takeuchi2014} estimated the signal from large-scale filaments using equilibrium ionization models with a  \citet{haardt2012} UV background. They found that this signal can be detected only with future instruments such as the \texttt{SKA}. However, their approach does not account for radiative transfer effects or a self-consistent source evolution. Our work improves on this by employing state-of-the-art cosmological hydrodynamical simulations post-processed with an accurate radiative transfer code. 


\citet{khullar2020} employed a numerical setup \citep[based on][]{Marius2018,Marius2020} similar to ours, but focused exclusively on the $z \geq 7$ Universe. They found that, at these early times, the peak of the $^{3}\mathrm{He}^{+}$ signal lies in the range $\sim 1$--50\,$\mu$K, and that in the vicinity of quasars the signal is consistently in emission. Their calculation assumes the spin temperature to be equal to the gas kinetic temperature, which likely provides an upper limit to the signal amplitude (mainly the power spectra). Based on these results, they concluded that it is difficult to distinguish between reionization histories driven by different source populations. This is likely a consequence of the fact that, at such early epochs, the evolution of He\,\textsc{ii} is predominantly governed by stellar sources and the impact of other types of sources are fairly negligible.
Recent work by \citet{Spina2026} suggests that the observable $^{3}\mathrm{He}^{+}$ signal becomes more sensitive to the underlying helium reionization history at later times, where quasars dominate the ionizing background. They show that the redshift evolution and amplitude of the corresponding power spectrum can differ significantly between reionization scenarios, indicating that the signal may retain distinguishable imprints of the reionization timeline even if its absolute amplitude remains weak.

Using the most recent helium reionization simulations \citep{Basu2024}, obtained by post-processing outputs from the hydrodynamical simulation \texttt{TNG300} \citep{volker2018,naiman2018,marinacci2018,pillepich2018,nelson2018} with the radiative transfer code \texttt{CRASH} \citep{ciardi2001,maselli2009,graziani2018,glatzle2022}, we examined the spatial and temporal evolution of the 3.5cm signal and its connection to the ionization state of helium driven by quasars. We also discussed relevant probes related to this signal to constrain the
helium reionization. Our main findings are as follows:

\begin{itemize}

    \item The spatial distribution of the differential brightness temperature, $\delta T_{\mathrm{b},^{3}\mathrm{He}^{+}}$, shows a clear transition from He\,\textsc{ii}-dominated gas at $z = 4$ to a He\,\textsc{iii}-dominated one at $z = 2.90$. The volume-averaged $\delta T_{\mathrm{b},^{3}\mathrm{He}^{+}}$ remains relatively constant until $z=3$, after which it declines rapidly from $\sim 10^{-2} \mu$K to $\sim 8 \times 10^{-5} \mu$K by $z = 2.3$ as reionization progresses, mirroring the evolution of the He\,\textsc{ii} fraction. The standard deviation increases slightly during the early stages of reionization and then decreases as the IGM becomes more (uniformly) ionized.

    \item The 3.5cm power spectrum, $\rm{\Delta_{^{3}He^{+}}}(\mathit{k})$, increases with \( k \), indicating stronger small scale fluctuations. Initially, these are driven by gas density and temperature variations. The simulated power spectra are 5-6 orders of magnitude below the current observational upper limits set by \citet{Trott2024}. Even under the assumption of full $T_{\rm S}$ coupling with the gas (which provides an upper bound on the simulated signal), the predicted amplitudes remain well below the observational limits. This highlights the challenge of detecting the signal with current instruments.

    \item The 3.5cm-AGN cross-power spectrum evolves from a large-scale positive correlation at $z = 4.18$, peaking at $k \sim 1\,h\,\mathrm{cMpc^{-1}}$, to a strong anti-correlation by $z = 3.28$ as He\,\textsc{iii} bubbles grow and suppress 3.5cm emission around AGNs. The peak of the spectrum shifts to larger spatial scales, reflecting typical bubble sizes, and by $z = 2.90$ the signal weakens further, with the correlation approaching zero on small scales.


\end{itemize}

Overall this study highlights the potential of the $^{3}\mathrm{He}^{+}$ hyperfine transition line as a probe of the latest stages of helium reionization driven by quasars. Although current observational limits lie well above the theoretical predictions, upcoming improvements in radio instrumentation are expected to narrow this gap, similar to the progress seen in 21cm studies.
To obtain more reliable insights, future progress will rely on improved instrumental sensitivity to achieve the high signal-to-noise ratios required for detecting the $^{3}\mathrm{He}^{+}$ signal (as also discussed by \citet{Spina2026}). Our work will help exploring instrumental noise and foreground modeling to improve our understanding of observational constraints. The enhanced sensitivity of the \texttt{SKA}, particularly \texttt{SKA-mid}, will be pivotal in improving the detectability of the $^3\mathrm{He}^+$ hyperfine signal and advancing our understanding of the final stages of helium reionization.

\section*{Acknowledgements}
All simulations were carried out on the machines of Max Planck Institute for Astrophysics (MPA) and Max Planck Computing and Data Facility (MPCDF). AB thanks the entire EoR research group of MPA for all the encouraging comments for this project. We thank the anonymous reviewer for useful comments which helped to improve the manuscript. We thank Saleem Zaroubi and Cathryn Trott for valuable discussions in the initial stage of the study. This work made extensive use of publicly available software packages : \texttt{numpy} \citep{vander2011}, \texttt{matplotlib} \citep{Hunter2007}, \texttt{scipy} \citep{Jones2001}, \texttt{tools21cm} \citep{Giri2020} and \texttt{CoReCon} \citep{Garaldi2023}. Authors thank the developers of these packages.

\section*{Data Availability}
The final data products from this study will be shared on reasonable request to the authors.



\bibliographystyle{mnras}
\bibliography{mnras} 




\appendix
\section{Impact of the Spin Temperature Treatment on 3.5cm Signal: Additional Results}
\label{compared_PS_with_Ts_treatment}

Here, we quantify the effect of the spin temperature ($T_{\rm S}$) modelling on the predicted cross-correlation between the 3.5cm signal and AGNs
. As discussed in Section~\ref{signal_def}, we compare two limiting cases for the computation of $T_{\rm S}$. To isolate the impact of the spin temperature assumption, all other simulation ingredients and post-processing steps are kept identical between the two cases.

\begin{figure}
    \includegraphics[width=\columnwidth]{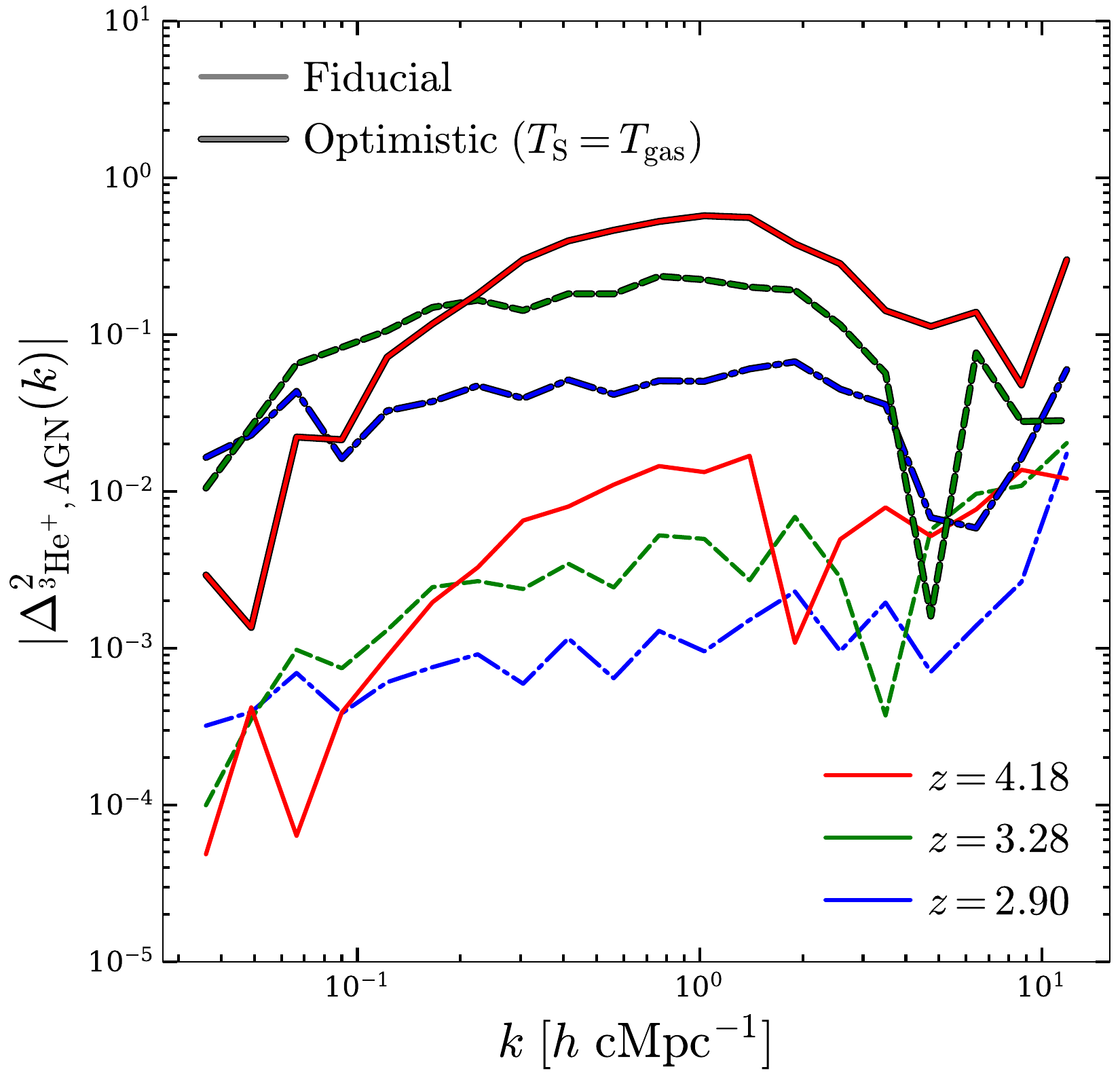}
    \caption{Comparison of the 3.5cm-AGN dimensionless cross-power spectra under different $T_{\rm S}$ assumptions. The fully coupled case ($T_{\rm S} = T_{\rm gas}$) is shown with black-edged curves, while the self consistent case follows the same redshift color coding as in the main text.}
    \label{3he_crossPS_compare}
\end{figure}

\begin{figure}
    \includegraphics[width=\columnwidth]{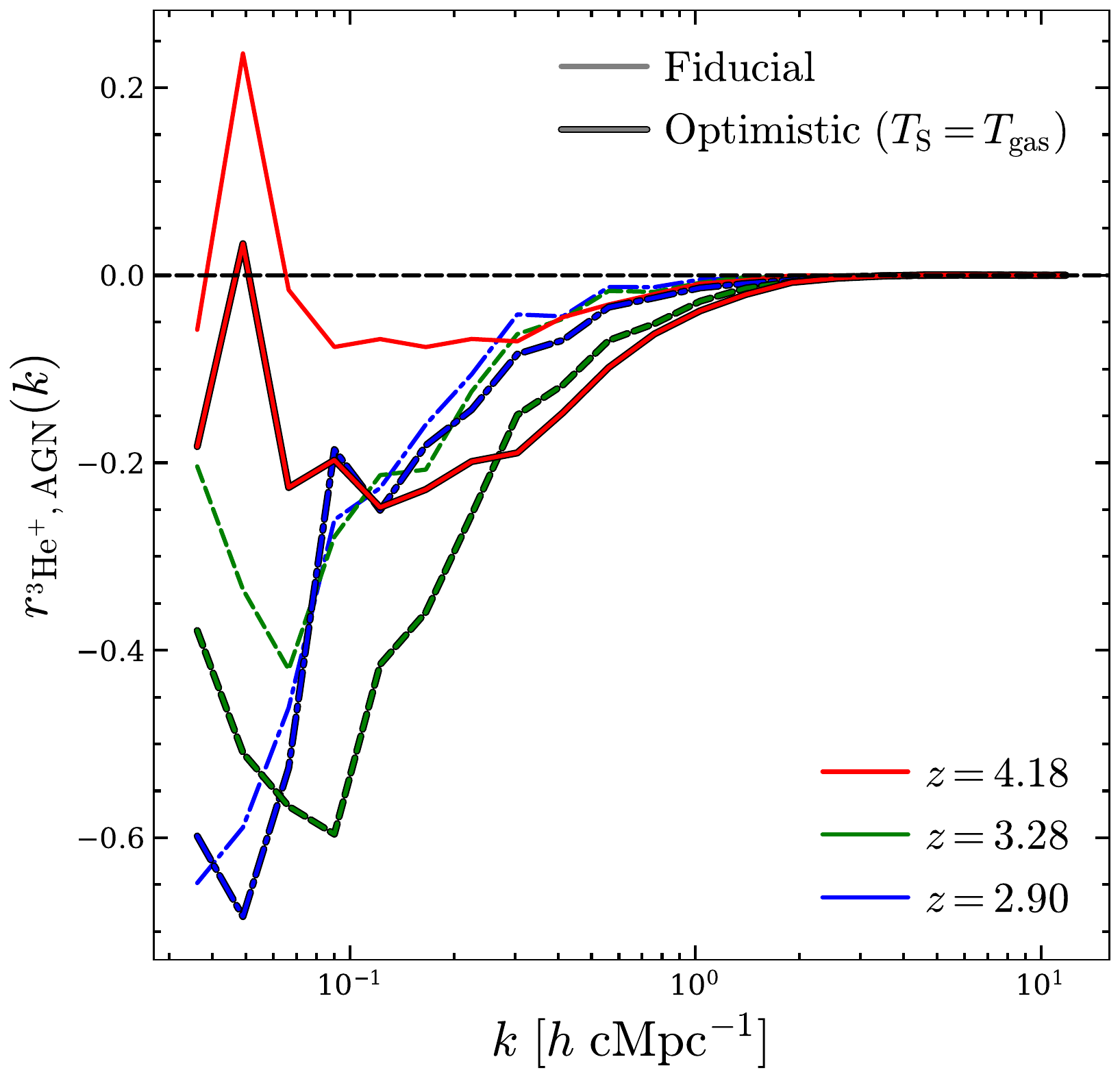}
    \caption{Comparison of the 3.5cm-AGN cross-correlation coefficient under different $T_{\rm S}$ assumptions. The fully coupled case ($T_{\rm S} = T_{\rm gas}$) is shown with black-edged curves, while the self consistent case follows the same redshift color coding as in the main text.}
    \label{3he_crossr_compare}
\end{figure}

In Figure~\ref{3he_crossPS_compare}, we compare the cross-correlation power spectra between the 3.5cm signal and AGNs. The overall behaviour is broadly similar to that seen in Figure~\ref{he3_qso_crossPS}. The amplitude is systematically higher in the fully coupled `optimistic' case than in the `fiducial' scenario, reflecting the stronger response of the brightness-temperature field to the underlying gas and source distribution. However, the scale- and redshift-dependent trends remain largely unchanged between the two treatments. In contrast, the cross-correlation coefficient is systematically lower in the fully coupled case, as shown in Figure~\ref{3he_crossr_compare}. This occurs because, although the signal amplitude increases, additional fluctuations arising from gas density and temperature variations are also enhanced, reducing the overall coherence between the 3.5cm signal and the AGN distribution.

Overall, the fully coupled `optimistic' limit provides an upper bound on the expected signal amplitude, whereas the self-consistent calculation yields a more realistic estimate. The true astrophysical signal is expected to lie between these two cases, depending primarily on the poorly constrained Ly$\alpha$ background and the efficiency of radiative coupling in the high-redshift intergalactic medium.



\bsp	
\label{lastpage}
\end{document}